\documentstyle[twocolumn,epsfig,prb,aps,amsmath,amssymb]{revtex}
\begin{document}

\twocolumn[\hsize\textwidth\columnwidth\hsize\csname
@twocolumnfalse\endcsname

\title{Antiferromagnetism in doped anisotropic two-dimensional
spin-Peierls systems}

\author{A.~Dobry$^{a,b}$, P.~Hansen$^b$, J.~Riera$^{a,b}$,
D.~Augier$^a$ and D.~Poilblanc$^{a,b}$}
\address{
$^a$Laboratoire de Physique Quantique \& Unit\'e Mixte
de Recherche CNRS 5626\\
Universit\'e Paul Sabatier, 31062 Toulouse, France\\
$^b$Instituto de F\'{\i}sica Rosario, Consejo Nacional de 
Investigaciones 
Cient\'{\i}ficas y T\'ecnicas y Departamento de F\'{\i}sica\\
Universidad Nacional de Rosario, Avenida Pellegrini 250, 2000-Rosario,
Argentina}
\date{\today}
\maketitle
\begin{abstract}

We study the formation of antiferromagnetic correlations 
induced by impurity doping in 
anisotropic two-dimensional spin-Peierls systems.
Using a mean-field approximation to deal with the inter-chain 
magnetic coupling, the intra-chain correlations are treated exactly
by numerical techniques.
The magnetic coupling between impurities is computed
for both adiabatic and dynamical lattices 
and is shown to have an alternating sign as a function of the
impurity-impurity
distance, hence suppressing magnetic frustration.
An effective model based on our numerical results
supports the coexistence of antiferromagnetism and dimerization
in this system.


\end{abstract}

\vskip2pc]

General interest for spin-Peierls (SP) systems was recently renewed
by the discovery of CuGeO$_3$~\cite{CuGeO}, the first inorganic SP
material. The SP transition is characterized by 
a freezing of the spin fluctuations below an energy scale given
by the spin gap $\Delta_S$ accompanied by a simultaneous lattice 
dimerization~\cite{structure}. 
Rich phase diagrams have been obtained experimentally upon doping
this compound with non-magnetic
impurities~\cite{Zn_doped,Mg_doped}.
In site-substituted systems such as (Cu$_{1-x}$M$_x$)GeO$_3$, where
M=Zn (Ref.~\onlinecite{Zn_doped}) or Mg (Ref.~\onlinecite{Mg_doped}),
long range antiferromagnetic (AF) order is stabilized at low
temperature while the dimerization still persists (D-AF phase). 
In Mg-doped compounds, for impurity concentrations larger than a
critical value ($x_c\simeq 0.02$), a first order transition occurs 
between the D-AF phase and a uniform AF (U-AF) phase where the
dimerization disappears.~\cite{Mg_doped}
The coexistence of the two types of order in the D-AF phase
is an intriguing phenomenon since lattice dimerization favors the
formation of spin singlets on the bonds while low energy spin
fluctuations exist in an AF phase.

Theoretically, the effect of impurity doping in SP systems was 
considered for fixed-dimerized~\cite{Laukamp},
adiabatic~\cite{Fukuyama,Khomskii,Hansen_imp} 
and quantum-dynamical~\cite{Hansen_imp} lattices. 
A single nonmagnetic impurity 
releases a soliton in the chain which can be viewed
as a kink in the lattice distortion. 
In the absence of interchain couplings,
such an excitation can freely propagate away
from the impurity. On the other hand, the interchain
elastic coupling $K_\perp$
was shown to produce confinement within some distance from the
impurity~\cite{Khomskii,Hansen_imp}.

For a finite impurity concentration, the coexistence between SP and
AF orders has been previously discussed
either considering randomly distributed domain walls in a $XX$ 
chain~\cite{Fabrizio} or assuming
small fluctuations of the magnetic exchange 
constants~\cite{Khomskii_2}. Despite their success to describe some
experimental results, these models are rather limited since they
do not take into account the microscopic origin of the
soliton formation nor the interchain couplings. 
In this paper, a realistic microscopic model with interchain 
magnetic and elastic couplings is considered to describe
the formation of a region with AF correlations in the vicinity
of each impurity and which allows an estimation of the effective
interaction between impurities in the two-dimensional (2D) system. 
Thus, we are able to construct and study an effective model in order
to understand the effects of a finite impurity 
concentration.~\cite{Kivelson}

In a first step, the spin-phonon coupling
is treated in the adiabatic approximation. The Hamiltonian
${\cal H}={\cal H}_{\mathrm{mag}} + {\cal H}_{\mathrm{el}}$ 
is:
\begin{eqnarray}
{\cal H}_{\mathrm{mag}}&=&J_\parallel\sum_{i,a} (1+\delta_{i,a})\,
{\bf S}_{i,a} \cdot {\bf S}_{i+1,a} + J_\perp
\sum_{i,\langle a,b\rangle} {\bf S}_{i,a} \cdot {\bf S}_{i,b} \ , 
\nonumber \\
{\cal H}_{\mathrm{el}}&=& \sum_{i,a}\{ \frac{1}{2} K_\parallel
\delta_{i,a}^2 + K_\perp\, \delta_{i,a}\delta_{i,a+1} \}\ ,
\label{hamad} 
\end{eqnarray}
\noindent
where $a$ is a chain index and $i$ labels the sites along the chains.
Atomic displacements are only considered along the chain direction,
$\delta_{i,a}$ being here a classical variable 
related to the change of the bond length 
between sites $(i,a)$ and $(i+1,a)$.
The magnetic part includes a magnetoelastic coupling 
$J_\parallel$ (hereafter set to unity)
and an exchange interaction $J_\perp$ connecting nearest neighbor (NN) 
chains. We eventually include in our model a next NN exchange
interaction along the chain
whose relevance for CuGeO$_3$ has been
emphasized~\cite{Riera-Dobry,Castilla}.
${\cal H}_{\mathrm{el}}$ is the elastic energy.
The interchain elastic interaction ($K_\perp$) is limited to NN chains.
Stability of the lattice implies $K_\parallel\ge 2 |K_\perp|$.
Typical values of the parameters for CuGeO$_3$ are $J_\perp\sim 0.1$ 
(Ref.~\onlinecite{Regnault_Jperp}) and 
$K_\perp/K_\parallel\sim 0.2$ (Ref.~\onlinecite{elastic_CuGeO}).

In order to study numerically model~(\ref{hamad}),
we treat exactly the single chain problem using exact
diagonalization (ED) or Quantum Monte Carlo (QMC) methods, while
the interchain magnetic coupling is treated in a self-consistent 
mean-field (MF) approximation. This is an standard procedure to 
include interchain couplings in the study of quasi-one-dimensional
systems.\cite{Schulz} 
Moreover, Inagaki and Fukuyama\cite{Inagaki} have used  a similar
MF approximation to treat the interchain coupling in the 
bosonized version of ~(\ref{hamad}) within
a self-consistent harmonic approximation.
Thus, in our procedure,    
the interchain magnetic coupling is replaced by its MF form:
\begin{eqnarray}
{\cal H}_{\mathrm{MF}}^\perp= J_\perp
\sum_{i,\langle a,b\rangle} \{\langle S_{i,a}^z\rangle S_{i,b}^z
+S_{i,a}^z \langle S_{i,b}^z\rangle
- \langle S_{i,a}^z\rangle \langle S_{i,b}^z\rangle\}.
\label{HMF}
\end{eqnarray}

By extending a similar approach previously applied to
one-dimensional (1D) chains~\cite{Feiguin,Hansen_imp} 
to the case of the 2D lattice, a sweep
is performed in the transverse direction, i.e. $a\rightarrow a+1$.
For each chain $a$, we compute the MF values
$\langle S_{i,a}^z\rangle$
and the classical variables $\{\delta_{i,a}\}$
by energy minimization, which is achieved by solving iteratively 
the equations
\begin{eqnarray}
\delta_{i,a}=
-\{ J_\parallel\langle {\bf S}_{i,a} \cdot {\bf S}_{i+1,a}\rangle
+K_\perp(\delta_{i,a+1}+\delta_{i,a-1})\}/ K_\parallel.
\label{eq} 
\end{eqnarray}
Then, these new values of the AF and SP order parameters
enter as input for the chain $a+1$.
This procedure is iterated until convergence is reached.
In this way, we can study numerically finite clusters consisting
of $N$ coupled chains with $L$ sites where, typically,
$N\times L=12\times 18$ in ED and $N\times L=6\times 40$ in QMC,
with toroidal boundary conditions.

A similar MF approach can be adapted to study a model
equivalent to (\ref{hamad}) but with quantum phonon degrees of freedom.
In this case, phonon operators $b_{i,a}^\dagger$ and $b_{i,a}$
are introduced on each bond and the displacements $\delta_{i,a}$ 
become $g(b_{i,a}^\dagger+b_{i,a})$,
where $g$ is the magnetoelastic constant. Then, the
classical elastic term ${\cal H}_{\mathrm{el}}$ is
replaced by its quantum version,
\begin{eqnarray}
{\cal H_{\mathrm{ph}}}=\Omega \sum_{i,a} \{ b_{i,a}^\dagger
b^{\phantom\dagger}_{i,a}
+ \Gamma (b^{\phantom\dagger}_{i,a} +
b_{i,a}^\dagger)
(b^{\phantom\dagger}_{i,a+1} + b_{i,a+1}^\dagger) \} \ ,
\label{hamdy} 
\end{eqnarray}
\noindent
where $\Gamma=K_{\perp}/(2K_{\parallel})$, and the phonon frequency 
$\Omega$ is related to $g$ by
$\Omega=2g^2K_\parallel$. The adiabatic limit (\ref{hamad}) is
recovered when $\Omega\rightarrow 0$ (requiring $g\rightarrow 0$
also). Similarly to the 
interchain magnetic term, the interchain
elastic term of (\ref{hamdy}) is then treated in mean-field by
introducing a lattice order parameter
$\delta_{i,a}^{\mathrm{MF}}=g\langle b_{i,a}^\dagger+b_{i,a}\rangle$.
Then, the term (\ref{hamdy}) is replaced by:
\begin{eqnarray}
{\cal H_{\mathrm{ph,MF}}}=\Omega \sum_{i,a} \{ b_{i,a}^\dagger
b^{\phantom\dagger}_{i,a}
+ \frac{\Gamma}{g} (b^{\phantom\dagger}_{i,a} +
b_{i,a}^\dagger)
\delta^{MF}_{i,a+1}  \} \ ,
\label{hamdyMF} 
\end{eqnarray}
\noindent
Note that in this case it is not necessary to solve an equation 
similar to (\ref{eq}).
To diagonalize the single chain Hamiltonian with $L\le 8$,
a Lanczos algorithm is used.
The phononic degrees of freedom are treated within 
a variational formalism previously introduced 
~\cite{Fehrenbacher,Augier_dyn,Hansen_imp}.
Note that inelastic neutron scattering experiments~\cite{Braden} on
CuGeO$_3$ reveal a rather large phonon 
frequency $\Omega/J_\parallel\simeq 2$ suggesting large lattice
quantum effects in this material.

\begin{figure} 
\begin{center}
\epsfig{file=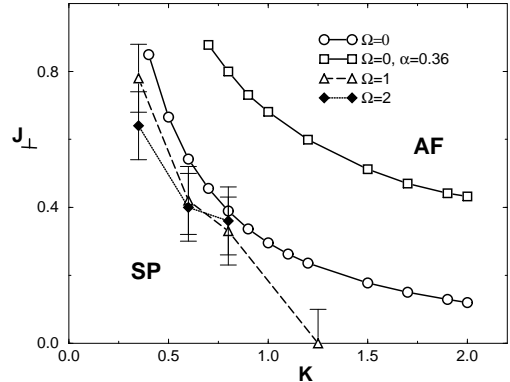,width=6cm,angle=-90}
\end{center}
\caption{Phase diagram of the pure system as a function of
$K=K_\parallel-2|K_\perp|$ and $J_\perp$.
The full lines correspond to the adiabatic results obtained by ED
for chains with up to $L=18$ sites, $\alpha=0$ (circles) and
$\alpha=0.36$ (squares).
The dashed (dotted) line corresponds to the calculation with
quantum phonons for $L=8$, $K_\perp=0.2$ $\alpha=0$,
and $\Omega=1$ ($\Omega=2$)}
\label{pure}
\end{figure}

As a preliminary study, we apply the MF procedure to 
the case of an homogeneous system without impurities.
In models~(\ref{hamad}) and ~(\ref{hamdy}),
$J_\perp$ is expected to stabilize the
AF state while a small $K_\parallel$ (or large
magnetoelastic coupling $g$) tends to favour SP order.
For each value of the couplings 
$J_{\perp}$ and $K$ 
(where
$K=K_\parallel-2|K_\perp|$ is the relevant parameter in the SP phase)
we obtain the ground state without imposing any restriction on the
MF parameters. 
We found only two different phases, the SP phase where
$\langle S_z \rangle=0$ and
$\delta_{i,a}\neq 0$ and the antiferromagnetic (AF) phase with
$\langle S_z\rangle \neq 0$ and  $\delta_{i,a}=0$.
Then, the phase diagram in the $K-J_{\perp}$ plane, 
can be obtained in a more efficient way by a direct comparison of the
energies of the N\'eel antiferromagnetic phase, 
and of the uniformly dimerized phase.
The phase diagram shown in Fig.~\ref{pure} exhibits a 
transition line between AF and SP phases. In the adiabatic case,
this line could be fitted by a law $J_{\perp}=\frac{A}{K}+B$
with $A=0.3656$ and $B=-0.06$ (this artificial small
negative value may be a consequence of small finite size effects,
see e.g., Ref. \onlinecite{Feiguin}).
A phase boundary with a form 
$J_{\perp}=\frac{A}{K}$ has been predicted by Inagaki and Fukuyama
\cite{Inagaki}.
However their bosonized approach do not fix unambiguously the value of 
$A$. 
 
In the case of the adiabatic calculation 
finite size effects were shown to be small for $K < 2$.
On the other hand, for the dynamical lattice, the calculation 
is reliable only for larger lattice couplings 
(i.e. smaller values of $K$)
due to stronger finite size effects. As expected, 
for very small $K$, 
lattice quantum fluctuations are less effective in dimerizing 
the chain than the adiabatic lattice. Then, the phase boundary
obtained with quantum phonons is located below the adiabatic one.  
This tendency becomes clear as $\Omega$ is increased, as shown in the 
figure. On the other hand, it has been suggested that  
dynamical phonons induce an effective magnetic frustration\cite{Uhrig}. 
This frustration which becomes relatively important for
larger $K$ destabilizes the AF phase thus moving the phase
boundary upwards as seen in the figure.
Consistently with this behaviour, if 
a next NN exchange term is included in the
Hamiltonian, in the {\em adiabatic} approximation, the SP phase is
more stable and the phase boundary is located above the corresponding
curve for $\alpha=0$. 
This behavior is shown in Fig.~\ref{pure} for the realistic
value of $\alpha=0.36$ obtained for CuGeO$_3$
(Ref. \onlinecite{Riera-Dobry}), where $\alpha$ is the
value of the next NN exchange coupling constant in units of
$J_{\parallel}$. We have checked that the set of realistic parameters
$K \approx 20$ and $K_\perp$, $J_\perp$ mentioned above corresponds
to a point in the SP phase.

\begin{figure}
\begin{center}
\epsfig{file=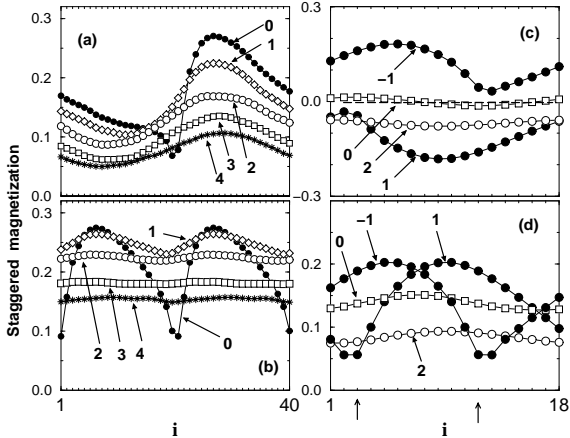,width=6cm,angle=-90}
\end{center}
\vspace{0.2cm}
\caption{
Typical patterns of the local staggered magnetization
$M^{\rm stag}_{i,a}$ along the chains. 
The transverse coordinate $a$ is indicated for each curve
(only a few chains of the 2D cluster are shown).
Closed circles correspond to the chains with an impurity.
(a-b) QMC results for a $8\times 40$ cluster with 
$K_\parallel=1.9$, $K_\perp=0.2$
and $J_\perp=0.1$; (a) single impurity case ($i_0=21$);
(b) 2 impurities on the same chain at a distance $\Delta i=L/2=20$
($i_1=1$ and $i_2=21$).
(c-d) ED results for two impurities located on next NN chains at a 
distance of $\Delta i=9$ along the chain (as indicated by the arrows)
on a $12\times 18$ cluster for $K_\parallel=2.4$, $K_\perp=0.2$
and $J_\perp=0.1$. Singlet (d) and triplet (c) 
configurations are shown.
}
\label{pattern}
\end{figure}

To start our analysis of impurity doping,
we consider a single impurity in order
to investigate the appearance of AF correlations in the SP phase.
As mentioned above, the impurity releases in the chain a topological
spin-1/2 solitonic excitation~\cite{Khomskii,Hansen_imp}
characterized by a change of parity of the dimerization order
which occurs in a finite region of longitudinal size $\xi_\parallel$
given by the soliton width.
The local magnetization on each chain $a$ can be decomposed into
uniform and staggered components~\cite{Eggert}, 
$\langle S_{i,a}^z\rangle=M_{i,a}^{\rm{unif}}+
(-1)^{i+a}M^{\rm stag}_{i,a}$. In fact, the excess 
uniform component $S^z_{\rm{sol}}=\pm\frac{1}{2}$ and the soliton,
characterized by a broad maximum of $M^{\rm stag}_{i,a}$,
remain confined in the chain with the impurity. However, as seen
in Fig.~\ref{pattern}(a),

\noindent
the interchain {\it magnetic}
coupling $J_\perp$ generates a large staggered component with the
same parity, i.e. $M^{\rm stag}_{i,a}$ keeping the same sign,
in the neighboring chains.
Simultaneously, the amplitude of the SP dimerization is 
significantly suppressed compared to the bulk value 
i.e. far away from the impurity. Large AF correlations can
be seen up to more than four chains away from
the impurity chain for magnetic couplings as small as $J_\perp=0.1$,
in particular
in the vicinity of the SP $\rightarrow$ AF transition line of
Fig.~\ref{pure}. 
The transverse range of the AF `polarization cloud' around the
impurity increases strongly with the transverse
coupling $J_\perp$.

A crucial feature of the polarization surrounding the impurity-soliton
area is that the sign of $M^{\rm stag}_{i,a}$
is unambiguously fixed by the orientation
($S^z_{\rm{sol}}=\pm\frac{1}{2}$)
of the soliton and by the position $(i_0,a_0)$ of the impurity
in such a way that $\rm{sign} \{ M^{\rm stag}_{i,a}\}
=\rm{sign}  \{ S^z_{\rm{sol}}\}\, (-1)^{a_0+i_0+1}$. 
This fact can be simply understood in the strong dimerization limit
($\delta\rightarrow 1$) where the introduction of the impurity on a
given site releases a spin-1/2 on one of its neighboring sites by
breaking a singlet bond. For smaller
lattice coupling, the excess spin can effectively hop from site
to site on the same sublattice (due to the underlying dimerization),
hence producing AF correlations with the parity defined above. 

\begin{figure}
\begin{center}
\epsfig{file=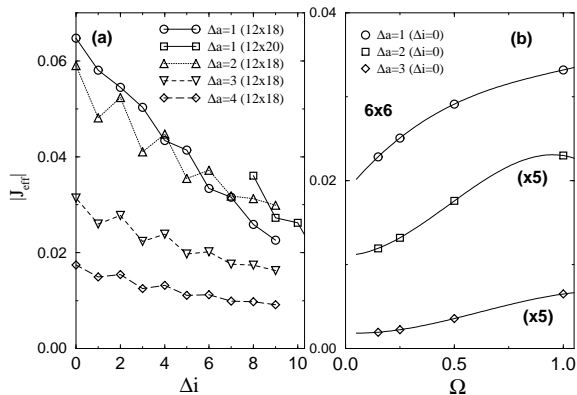,width=6cm,angle=-90}
\end{center}
\vspace{0.2cm}
\caption{
Absolute value of the effective magnetic coupling 
between two impurities located on different
chains separated by $\Delta a$; (a) vs. the 
separation $\Delta i=i_2-i_1$ along the
chains for an adiabatic lattice ($\Omega=0$) and for
$K_\parallel=2.4$, $K_\perp=0.2$ and $J_\perp=0.1$; (b) vs. the frequency 
$\Omega$ for $K_\parallel=1$, $K_\perp=0.2$ and $J_\perp=0.1$.
The sizes $N\times L$ of the clusters used in the calculations 
are indicated on the plots.
}
\label{jeff_fig}
\end{figure}

Let us now consider two impurities introduced 
simultaneously on two sites $(i_1,a_1)$ and $(i_2,a_2)$
of different chains ($a_1\ne a_2$). When the two polarization clouds
associated to each soliton-impurity `pair' start to overlap, one expects
their interaction to depend on the relative orientation
of the two solitons. As seen in
Figs.~\ref{pattern}(c,d), quite different patterns correspond to
the singlet and triplet arrangements of the two spin-1/2 solitons. 
As confirmed by our calculations, the lowest energy is always obtained
for a spin state which leads to the {\it same} parity of the
staggered magnetization associated to each impurity, i.e. which avoids 
completely magnetic frustration. The simple argument
developed above for a single impurity then suggests that 
a triplet $S=1$ (singlet $S=0$) configuration is
favored when the two impurities are located on the same sublattice
(opposite sublattices). 
It is then appropriate to define an effective magnetic coupling
between the AF clouds associated to each impurity
by $J_{\rm eff}=E_{\rm{S=1}}-E_{\rm{S=0}}$.
For a wide range of parameters 
leading to a SP state in the bulk, we have numerically found that
$J_{\rm eff}$ is ferromagnetic (F) if the two impurities belong to
the same sublattice and antiferromagnetic in the opposite case.
This implies that the coupling between the two local moments
associated to the impurities is either F or
AF in such a way that no frustration occurs.
The magnetic coupling, for physical values of the parameters, 
can be fairly extended in space as seen in Fig.~\ref{jeff_fig}(a).
Its range is directly controlled by the overlap of the
polarization clouds.
 It follows roughly a behavior like
\begin{eqnarray}
J_{\rm{eff}} \approx J_0 (-1)^{\Delta a +\Delta i+1}
\exp{(-C_\perp\frac{\Delta a}{\xi_\perp})}
\exp{(-C_\parallel\frac{\Delta i}{\xi_\parallel})},
\label{jeff_eqn}
\end{eqnarray}
where $J_0 \sim J_{\perp}$, $C_\alpha$ are of order unity 
and $\xi_\alpha\sim J_\alpha/\Delta_S$.

To get an insight on how finite size effects might affect
our results we have increased our cluster size ($12 \times 18$)
in both directions.
The change of the transversal dimension has very little
effect because the polarization clouds are almost independent 
when the impurities are separated by more than four chains 
(see the very small values of $J_{\rm eff}$ for $\Delta a=4$ in 
Fig.~\ref{jeff_fig}(a)). On the other hand, by
increasing the length of each chain, some changes occur but
only when the impurities are located at the largest distances. 
However, the exponential decay of the effective interactions
with distance 
(see below) is not qualitative changed as it can be seen in 
~\ref{jeff_fig}(a)).
Therefore, we  expect that the numerical values of the {\em fitting 
parameters} would not be much affected by finite size effects 
leaving the overall behaviour essentially unchanged.
In particular, we believe that the presence of
long range AF order in the effective model (see discussion below) is
a robust feature not affected by finite size effects.

When two impurities are introduced on the same chain ($a_1=a_2$) 
two cases have to be distinguished.
If the impurities are located on the {\it same} sublattice
a similar behavior is observed as described above
(compare Fig.~\ref{pattern}(b) and Fig.~\ref{pattern}(d)), i.e
the effective interaction is {\it ferro}magnetic. However, the
magnitude of $|J_{eff}|$ is $\approx 0.4$ (very slowly decaying
as $\Delta i$ increases) for the parameters
of Fig.~\ref{jeff_fig}(a), i.e. much larger than the values
corresponding to impurities in different chains.
If the impurities belong to {\it different} sublattices then a chain
with even number of sites is cut into two segments with even number
of sites each. 
In the lowest energy configuration ($S=0$) no soliton-antisoliton
pair was observed for separations $\Delta i$ up to $20$ in agreement
with previous work~\cite{Hansen_imp}
and the triplet excitation energy remains large ($\sim \Delta_S$).
Then, one can expect that for larger chains, when the formation of
soliton becomes favourable, the effective interaction
between them will be AF and their magnitude will be of the order
of $\Delta_S$. In summary, one can assume
that the effective interaction between impurities on the same chain
has the same form of Eq.~(\ref{jeff_eqn}) with $\Delta a=0$ and
$J_0$ being now $\approx \Delta_S$. This form is similar to the
one adopted in Ref.~\onlinecite{Fabrizio} for impurities in a single
chain except that these authors do not consider the sublattice sign
alternation. Nevertheless, it should be noticed that this AF 
interaction should also decay for very large $\Delta i$.

When lattice quantum fluctuations are introduced ($\Omega > 0$),
the qualitative properties of the effective interaction
$J_{\rm{eff}}$ are preserved.
Consistently with the relatively larger stability of the AF phase
in the small $K$ region
shown in Fig.~\ref{pure} with respect to the adiabatic case, we have
found that lattice dynamics lead to an increase of the size of the AF
cloud associated to each soliton.
Therefore, the magnitude of the magnetic coupling $J_{\rm{eff}}$ 
increases with the phonon frequency $\Omega$ as shown
in Fig.~\ref{jeff_fig}(b). 

The final part of our study is the analysis of a simple effective
two-dimensional
spin-1/2 Heisenberg model between impurities with a long range
interaction given by Eq.~(\ref{jeff_eqn}). The `bare' parameters
are the same as in Fig.~\ref{jeff_fig} and the parameters of
the expression (\ref{jeff_eqn}) have been obtained by fitting the
curves shown in that figure and similar data for the case of
$\Delta a=0$. In the direction perpendicular to the chains we have
neglected the effective interactions beyond a distance 
$\Delta a =5$.
We have also assumed that even segments are associated to a
soliton-antisoliton pair.\cite{note}

\begin{figure}
\begin{center}
\epsfig{file=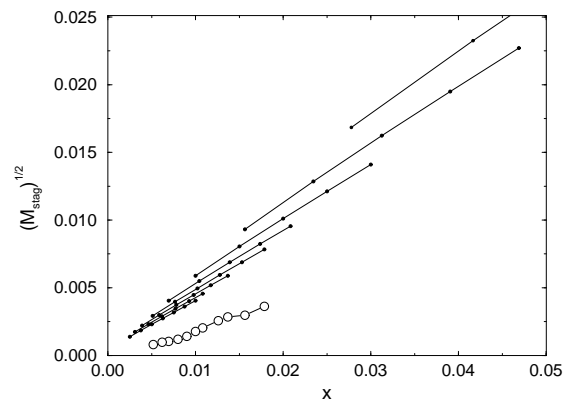,width=6cm,angle=-90}
\end{center}
\vspace{0.2cm}
\caption{Square root of the staggered magnetization (defined in
the text) as a function of the impurity doping for several 2D
clusters. Full circles correspond to, from top to bottom, $12 \times 12$,
$16 \times 16~$,$\cdots$, $40 \times 40$ clusters. The open circles
are the extrapolations to the bulk limit.
}
\label{effmodel}
\end{figure}

A given number of spin-1/2 impurities $4 \le N_{\rm imp} \le 16$
is thrown at random on systems of coupled chains of sizes up to
$40 \times 40$. Then, the staggered magnetization
$M_{\rm stag} = (1/N^2_s)(\sum_{i,a} (-1)^{i+a} S^z_{i,a})^2$, where
$N_s=N \times L$, is computed
and averaged over, typically, 12,000-16,000 random realizations.
The square root of this quantity is shown in Fig.~\ref{effmodel}.
By extrapolating to the bulk limit for a fixed impurity doping using
a polynomial in (1/$\sqrt{N_s}$) we found that $M_{\rm stag}$ is 
finite, implying long range AF order,
and slowly decreasing as $x$ goes to zero.
This behaviour is consistent with experimental
results\cite{Mg_doped} suggesting that
$M_{\rm stag}$ decays exponentially to zero as $x \rightarrow 0$.

In conclusion, spin-1/2 solitons released in 2D anisotropic SP systems 
by the introduction of impurities were shown to experience spatially
extended
F or AF exchange interactions depending on their relative positions.
These exchange interactions coexisting with the SP order are
calculated from realistic microscopic models and used to construct a
simple effective model which in turn enables us to show
the establishment of long range AF order and  to compute the AF order
parameter as a function of the impurity doping.

This work was supported in part by the ECOS-SECyT A97E05 program. 
We thank IDRIS (Orsay, France) and Florida State University for using
their supercomputer facilities.


\begin{references}
\bibitem{note} Additional simplifications have been
      adopted, specially for the case where more than two impurities
      fall on the same chain.
\end{references}
\end{document}